\documentstyle[preprint,aps]{revtex}
\begin{document}
\preprint{EHU-FT/9707}
\draft
\title{Damping rate of plasmons and photons \\ 
    in a degenerate nonrelativistic plasma} 
\author{M. A. Valle Basagoiti \thanks{\tt wtpvabam@lg.ehu.es}}
\address{Departamento de F\'\i sica Te\'orica, \\ 
Universidad del  Pa\'\i s Vasco, Apartado 644, E-48080 Bilbao, Spain}
\date{October 3, 1997}
\maketitle
\setcounter{page}{0}
\thispagestyle{empty}
\begin{abstract}
A calculation is presented of the plasmon and photon damping rates in 
a dense nonrelativistic plasma at zero temperature, 
following the resummation program of 
Braaten-Pisarski. At small soft momentum $k$, the damping is dominated 
by $3 \rightarrow 2$ scattering processes corresponding to double 
longitudinal Landau damping. The dampings are proportional to
 $(\alpha/v_{F})^{3/2}\, k^2/m$, where $v_{F}$ is the Fermi velocity.       
\end{abstract}


Over the last few years, there has emerged a systematic perturbation scheme
developed mainly by Braaten and Pisarski \cite{bra} and by 
Frenkel and Taylor \cite{fre}, which has finally allowed
the calculation of some thermodynamic and dissipative properties of 
quasiparticles in
ultra-relativistic plasmas (for a recent overview, see  \cite{review}). 
It is based on a resummation 
method in which the distinction between soft scales of order 
$g T$ or smaller and hard scales of order $T$ is completely crucial. 
Basically, when the quantities to be computed depend on soft scales,
the resummed perturbative expansion makes use of effective 
propagators and vertices accounting for long-distance medium effects as 
such Debye screening and Landau damping. The independence on the 
choice of gauge is guaranteed by the general properties of the hard 
thermal loops entering into the effective propagators and vertices 
\cite{bra,fre}   

One of the greatest success of these calculations has been the 
consistent calculation of the gluon damping rate at zero momentum 
\cite{gluon}, thus solving the plasmon puzzle in perturbative QCD at 
high temperature. Also, the fermion damping rate for an excitation at 
rest has been computed \cite{kob}. 
Other interesting quantities as
the damping rate of hard photons in the quark-gluon plasma \cite{kap}, and the 
lifetime of a moving fermion in QED plasmas also have been considered within 
this framework \cite{rebhan}. Only very recently, it has been shown by 
Blaizot and Iancu that 
the infrared divergences which plague this last quantity can be eliminated by 
a resummation based on the Bloch-Nordsieck mechanism at finite temperature 
\cite{blai}. 
All of these developments refer to a ultrarelativistic, 
high-temperature, zero-density regime. However, although degenerate plasmas 
(relativistic or not) are relevant in 
a wide area of applications ranging from metals to heavy ion collisions, 
relatively little work has been devoted to the computation of the damping 
of their excitations within the new developments \cite{olli,cris1}. In both 
of these references, 
the authors compute the damping of hard relativistic modes of 
momentum $k \sim k_F$, where $k_F$ is the Fermi momentum. 

Also, it is interesting to consider the problem of the 
damping rate of soft modes of momentum and energy of order 
$g k_{F}$ and $g E_{F}$ respectively. They correspond to 
collective excitations in the plasma. Actually, the perturbative
calculation of damping rates of these quasiparticles in the 
nonrelativistic regime is an old question. 
Over the sixties, there was some controversy concerning the damping 
rate of plasmons. A preliminar calculation was carried out by 
DuBois \cite{preli} who neglected screening effects. Ninham, Powell and
Swanson \cite{nih} repeated the calculation by including screening but missing 
out effective vertices which contribute 
to the same order in $e$ as the tree-level 
diagrams. Lately, DuBois and Kivelson \cite{dubois} corrected that  
procedure by taking into account the screening and
the effective  vertices and this seemed to be the final  
perturbative answer. However, the consistency of their method 
is not guaranteed because the mixing of the soft scales $(g E_{F}, g k_{F})$ and 
the hard scales $(E_{F}, k_{F})$ which is present in their calculations. 
Moreover, these authors do not report the numerical values of dampings.   

In this paper, I will largely follow the resummation program of  
Braaten and Pisarski to explicitly compute the plasmon and the photon 
damping rates at nearly zero momentum 
in a non relativistic degenerate electron gas 
to leading order in the electromagnetic fine structure constant
$\alpha = g_{e}^2/4\pi\simeq 1/137$. In this context, 
it is usual to introduce another 
coupling constant $r_{s} = (\alpha/v_{F}) (9 \pi/4)^{1/3}$. The 
degenerate limit is then $r_s \rightarrow 0$. 
Resummation  
is required in this limit because the plasma frequency
$\Omega_{p}= g_{e} k_F \sqrt{v_F/3}/\pi$ is soft versus the Fermi energy 
$E_F = k_{F}^2/2 m$, their ratio, of order $r_{s}^{1/2}$, 
playing the role of the coupling constant $g$ in the Braaten-Pisarski 
terminology.     

The main results of this paper are the formulae for dampings rates 
listed below in Eqs.~(\ref{fin1})--(\ref{fin3}).
The calculation, which is exact at leading order in $r_s$, does not include 
the effects of current-density and current-current interactions which lie
outside the non relativistic approximation and are suppresed by powers 
of the Fermi velocity over $c$. However, it is would be noted that 
the non-leading terms can be computed also within the framework 
described below.    

\begin{figure}
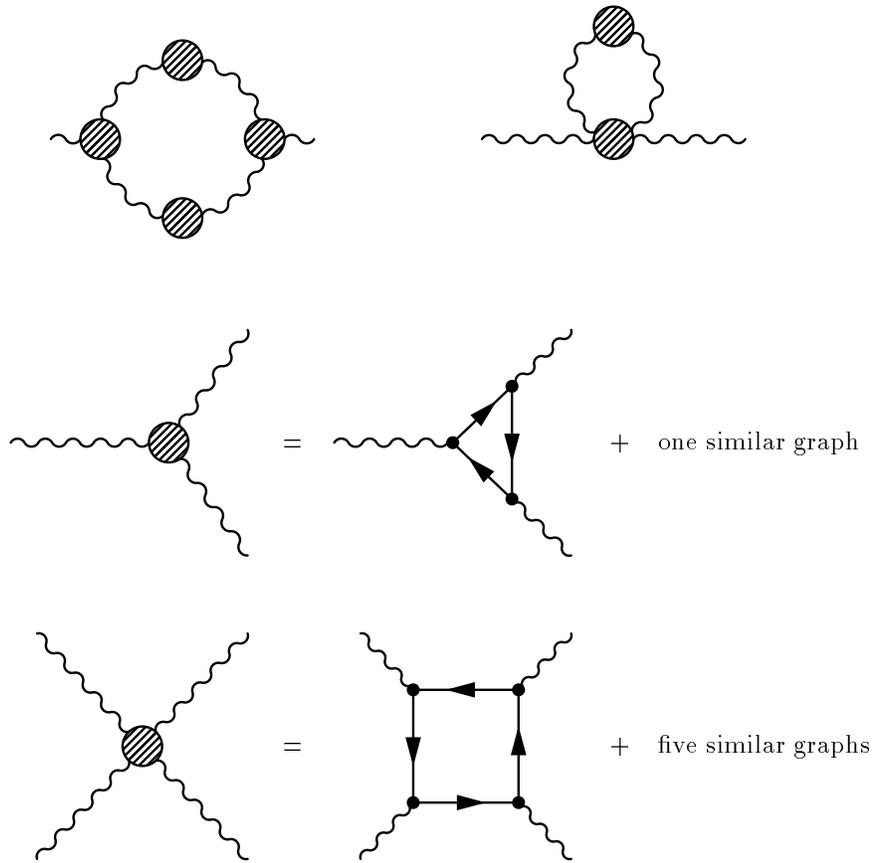

\caption{One-loop self-energy diagrams and effective vertices.}
\label{fig1}
\end{figure}

\begin{figure}
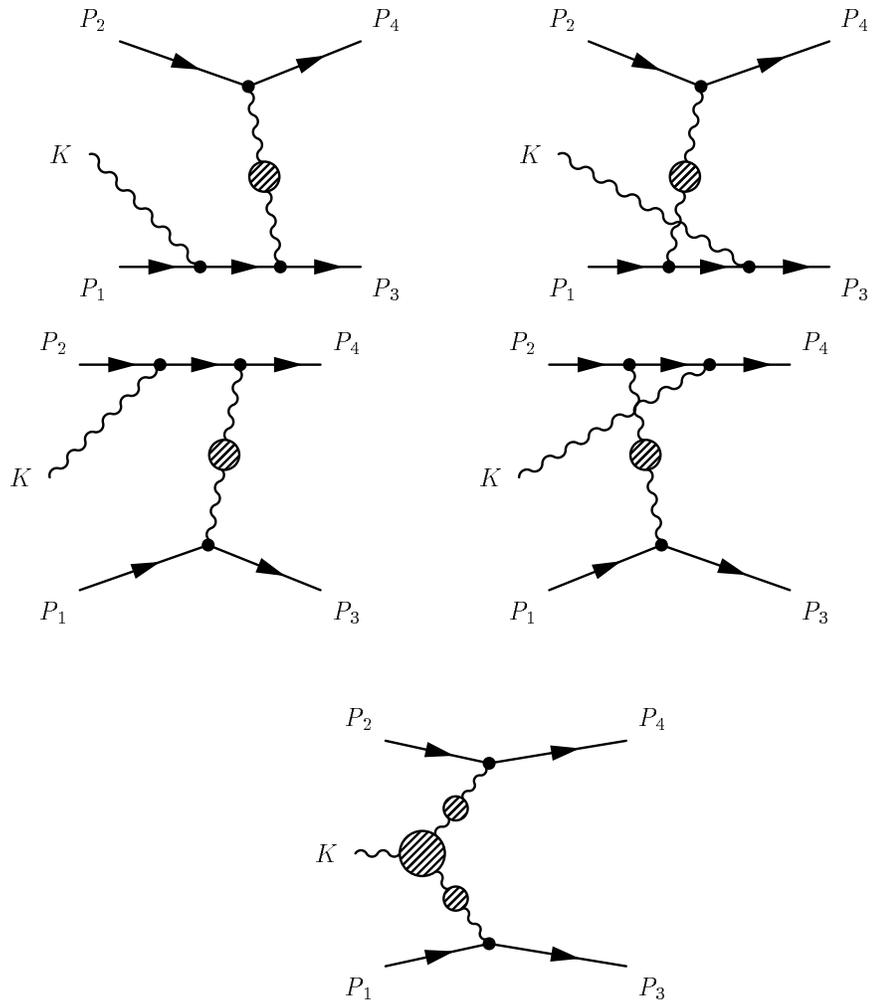

\caption{Direct scattering graphs for 
$\gamma^\ast(k) \,\mbox{e}(p_{1}) \, \mbox{e}(p_{2}) 
\rightarrow 
\mbox{e}(p_{3}) \, \mbox{e}(p_{4})$\,.}
\label{fig2}
\end{figure}

Let us proceed to outline our calculations. The scattering 
processes in the plasma contributing to the plasmon and photon damping 
rate to lowest order in $r_{s}$ can be identified by cutting the 
one-loop self-energy graphs with effective propagators and vertices 
drawn in Fig.~\ \ref{fig1}. There are three types of contributions depending on 
which piece of the imaginary part of effective propagators is picked. 
The pole-pole term corresponds to the decay of the plasmon or photon 
in a pair of plasmons which is forbidden by kinematics. The pole-cut 
term describes $2\rightarrow 2$ contributions which are zero since 
$k^0=\Omega_{p}$ is at or below threshold for these processes. The 
only nonzero contribution is the cut-cut. It comes from $3\rightarrow 
2$ processes: $\gamma^\ast(k) \,\mbox{e}(p_{1}) \, \mbox{e}(p_{2}) 
\rightarrow 
\mbox{e}(p_{3}) \, \mbox{e}(p_{4})$, two electrons from the Fermi sea 
end up above the Fermi surface through absorption of the plasmon. The damping 
arises enterely from the double Landau damping in the longitudinal 
exchange between the electrons outside and inside the Fermi sea as 
can be seen in Fig.~\ \ref{fig2}. 
It should be noted that the first four scattering 
diagrams 
arise from cuts in both   
effective three-vertices and effective four-vertex. The remaining 
five diagrams obtained by exchange of particles $P_{3} \leftrightarrow 
P_{4}$ do not enter to lowest order in $r_{s}$, since they come from 
cuts of higher-loop self-energy graphs.       

The damping rate $\Gamma_{l,t}(k)$ corresponding to this process is     
\begin{eqnarray}
\label{damp}
	\Gamma_{l,t}(k) & = &\frac{1}{2\omega_{l,t}(k)}\frac{1}{2! 2!} 
	   \int 
	   \frac{d^3 p_{1}}{(2 \pi)^3}\frac{d^3 p_{2}}{(2 \pi)^3}
	   \frac{d^3 p_{3}}{(2 \pi)^3}\frac{d^3 p_{4}}{(2 \pi)^3} 
	      n_{F}(p_{1}) n_{F}(p_{2}) 
	      [1 - n_{F}(p_{3})] [1 - n_{F}(p_{4})] \nonumber \\
	  &  &\times (2 \pi)^4 \delta(\omega_{l,t}(k) + E_{1} + E_{2} - 
	                              E_{3} - E_{4})
	               \delta^{(3)}({\bf k} + {\bf p}_{1} + {\bf p}_{2} -  
	                        {\bf p}_{3} - {\bf p}_{4})\,
	      \overline{|{\cal M}_{l,t}|^2}\,,
\end{eqnarray}
where $\overline{|{\cal M}_{l,t}|^2}$ denotes the scattering amplitude 
squared, averaged over the helicity states of the incoming photon and 
summed over the spin states of the initial and final electrons. A 
double factor $1/2!$ is included corresponding to the two pairs of 
identical particles present in the initial and final states. The 
electron energies are $E_{i}= p_{i}^2/2 m$ and the Fermi distribution 
is $n_{F}(p_{i})=\theta(k_{F}-p_{i})$. 
Now it is easy to write the matrix element
from the diagrams in Fig.2,
\begin{eqnarray}
\label{amp}
  {\cal M}_{l} & = & g_{e}^3 [g_{e}^2 \Pi^{(3)}(P_{3}- P_{1},P_{4}- P_{2}) 
                      D(\omega, {\bf p}_{3} - {\bf p}_{1})  
                      D(\omega^\prime ,{\bf p}_{4} - {\bf p}_{2}) + 
                     {\cal J}_{0} ] \epsilon_{0}(k)\,, \\
  {\cal M}_{t}(\lambda) & = & g_{e}^3 [ g_{e}^2 
  \Pi^{(3)}_{j}(P_{3}- P_{1},P_{4}- P_{2})
                      D(\omega, {\bf p}_{3} - {\bf p}_{1})  
                      D(\omega^\prime ,{\bf p}_{4} - {\bf p}_{2}) 
                       + 
                      {\cal J}_{j} ] \epsilon_{j}({\bf k},\lambda)\,.
\end{eqnarray}
The alert reader may wonder about the different 
orders of perturbation theory which seem to be mixed in the previous 
equations. In fact, we shall see below that the scattering amplitudes 
are of order $g_{e}^3$ due to cancellations between the effective 
vertices and propagators.                              

In the Coulomb gauge, the longitudinal effective propagator is 
\begin{equation}
\label{prop}
   D(\omega, k) = \frac{v_{F}^2}{v_{F}^2 k^2 - 3 \Omega_{p}^2\, 
   Q_{1}(\omega/v_{F} k)},
\end{equation}
where $Q_{1}$ is a Legendre function of the second kind,
\begin{equation}
 Q_{1}(x) = -1 + \frac{x}{2} \ln \left|\frac{1+x}{1-x}\right | - 
          i \frac{|x|}{2} \theta(1 - |x|). 
\end{equation}
This is not the retarded propagator but coincides with it for 
$\omega > 0$.      
Since the Fermi numbers in Eq.~(\ref{damp}) and the energy conservation
condition $\omega_{l,t}(k)= \omega  + \omega^\prime$ ensure that energy 
transfer variables 
$\omega = E_{3}- E_{1}$ and $\omega^\prime  = E_{4}- E_{2}$ are both 
positive, the effective vertices appearing in the amplitudes can be 
computed from their euclidean counterparts 
depending on imaginary 
frequencies. The complete calculation of the $\Pi$'s is rather involved, 
but fortunately only the `hard dense loop' part of them is required. 
In this approximation, one considers $\omega,\omega^\prime, p, k \ll 
E_{F}, k_{F}$. Besides, the leading contribution to the $\Pi$'s comes 
from a region in phase space where all internal fermionic momenta are 
in a narrow shell around the Fermi sphere, 
so that  
one can make an expansion of the integrand around the Fermi momentum 
after the imaginary frequency 
integration is performed.  Then, 
the analytic continuation with the retarded prescription 
from the imaginary 
frequencies to $\omega + i \varepsilon$ and 
$\omega^\prime + i \varepsilon$ gives 
\begin{eqnarray}\label{ver}
\Pi^{(3)}(\omega,\omega^\prime, {\bf p}, {\bf k}) & = &
  \frac{k_{F}\,{\bf k}\cdot {\bf p} }{\pi^2 \omega_{l}^2} \left[ 
       -Q_{1}(\omega/v_{F} p) + Q_{1}(\omega^\prime/v_{F} p) \right] 
       \nonumber \\ 
       & & +
 \frac{k_{F}\,{\bf k}\cdot {\bf p}^2}{2 \pi^{2} p^2 \omega_{l}^2} \left[ 
      \omega \frac{\partial}{\partial \omega} Q_{1}(\omega/v_{F} p) + 
      \omega^\prime \frac{\partial}{\partial \omega^\prime} 
       Q_{1}(\omega^\prime/v_{F} p) \right] \nonumber \\
       &  & - 
 \frac{2 \, k_{F}\,{\bf k}\cdot {\bf p}^2}{\pi^{2} p^2 \omega_{l}^3} \left[     
      \omega Q_{1}(\omega/v_{F} p) + 
                    \omega^\prime  Q_{1}(\omega^\prime/v_{F})\right] 
                    \nonumber \\
        & & +
 \frac{k_{F} k^2 p^2}{2 \pi^{2} p^2 \omega_{l}^2} \left[
        Q_{1}(\omega/v_{F} p) + 
        Q_{1}(\omega^\prime/v_{F} p) \right] + {\cal O}(k^3)\, , 
 \end{eqnarray}
 and 
 \begin{eqnarray}\label{verj}
 \Pi^{(3)}_{j}(\omega,\omega^\prime, {\bf p}, {\bf k}) & = &
  \frac{k_{F} \, p_{j}}{\pi^2 \omega_{t}} \left[ 
       -Q_{1}(\omega/v_{F} p) + Q_{1}(\omega^\prime/v_{F} p) \right] 
       \nonumber \\ 
       & & +
  \frac{k_{F}\,{\bf k}\cdot {\bf p}\, p_{j}}{2 \pi^{2} p^2 \omega_{t}} \left[ 
      \omega \frac{\partial}{\partial \omega} Q_{1}(\omega/v_{F} p) + 
      \omega^\prime \frac{\partial}{\partial \omega^\prime} 
       Q_{1}(\omega^\prime/v_{F} p) \right] \nonumber \\
       &  & -
  \frac{2 \, k_{F}\,{\bf k}\cdot {\bf p}\, p_{j}}{\pi^{2} p^2 
            \omega_{t}^2} \left[     
      \omega Q_{1}(\omega/v_{F} p) + 
                    \omega^\prime  Q_{1}(\omega^\prime/v_{F})\right] + 
                    {\cal O}(k^2)\, , 
 \end{eqnarray}
 where ${\bf p} = 1/2\, ({\bf p}_{3}-{\bf p}_{1}+{\bf p}_{2}-{\bf 
 p}_{4})$. Terms proportional to $k_{j}$ do not appear in 
 Eq.~(\ref{verj}) since they do not 
 contribute because the polarization vectors satisfy the Coulomb gauge 
 constraint ${\bf k}\cdot {\bf\epsilon}(\lambda) = 0$.
 
 The terms ${\cal J}_{0}$ and ${\cal J}_{j}$ in the amplitudes are 
 easily written from the real part of the non-relativistic electron 
 propagator because the virtual fermion is not on shell. They are of 
 the form 
 \begin{eqnarray}
 {\cal J}_{0}&=& D(\omega, {\bf p}_{3} - {\bf p}_{1}) \frac{1}
           {E_({\bf p}_{2})+\omega_{l} - E({\bf p}_{2}+ {\bf k})} + 
            \mbox{three similar terms}\, , \\ 
 {\cal J}_{j}&=& D(\omega, {\bf p}_{3} - {\bf p}_{1}) 
      \frac{(2 p_{2}^j + k^j) (2 m)^{-1}}
       {E_({\bf p}_{2})+\omega_{t} - E({\bf p}_{2}+ {\bf k})} + 
       \mbox{three similar terms}\, .  
 \end{eqnarray}
 
 Putting this together, we find the amplitudes to lowest order in $k$
 \begin{eqnarray}
 {\cal M}_{l} &=& g_{e}^3 \left[ \frac{k^2 p^2 - 4\,{\bf k}\cdot{\bf p}^2}
                        {m \omega_{l}^2} D(\omega,p) 
                         D(\omega^\prime,p) \right. \nonumber \\ 
              & &\;\;\;\;\;
              +\frac{ {\bf k}\cdot{\bf p}\,(2\,{\bf k}\cdot{\bf p}\, 
              m \omega^\prime + 2 \,{\bf p}_{2}\cdot{\bf k}\,p^2 - 
              {\bf k}\cdot{\bf p}\, p^2)}{m^2 p^2 \omega_{l}^3}
                D(\omega,p) \nonumber \\ 
              & &\;\;\;\;\; \left. 
              +\frac{ {\bf k}\cdot{\bf p}\,(2\,{\bf k}\cdot{\bf p}\, 
              m \omega - 2 \,{\bf p}_{1}\cdot{\bf k}\, p^2 - 
              {\bf k}\cdot{\bf p}\, p^2)}{m^2 p^2 \omega_{l}^3}
                D(\omega^\prime,p) \right] \epsilon_{0}(k) \, , \\
 {\cal M}_{t}(\lambda) &=& g_{e}^3 \left[
                         -\frac{4\,{\bf k}\cdot{\bf p}^2\, p^i}
                          {m \omega_{t}} D(\omega,p) 
                           D(\omega^\prime,p) \right. \nonumber \\ 
              & &\;\;\;\;\;
              +\frac{2\,{\bf k}\cdot \hat{{\bf p}}\,m \omega^\prime\, 
                            \hat{p}^i +
             {\bf k}\cdot{\bf p}\, p_{2}^i - 
                     {\bf k}\cdot{\bf p} \,p^i +
                     {\bf k}\cdot{\bf p}_{2}\, p^i}
                     {m \omega_{t}^2} D(\omega,p)  \nonumber \\ 
               & &\;\;\;\;\; \left.
              +\frac{2\,{\bf k}\cdot \hat{{\bf p}}\,m \omega\, 
                            \hat{p}^i-
                         {\bf k}\cdot{\bf p}\, p_{1}^i - 
                     {\bf k}\cdot{\bf p} \,p^i -
                     {\bf k}\cdot{\bf p}_{1}\, p^i }
                     {m \omega_{t}^2} D(\omega^\prime,p) \right]
                     \epsilon_{i}({\bf k},\lambda)\,.    
 \end{eqnarray}
 Because of simplifications between Legendre functions, these 
 expressions are far simpler that one might have expected given 
 the complicated form for the effective vertices.
 It should be noted the 
 cancellation  of terms of first and zeroth order in $k$ in 
 ${\cal M}_{l}$ and  ${\cal M}_{t}$ respectively. Also, they agree 
 with those previously computed, long time ago, in 
 Ref.\ \cite{dubois}. The terms 
 proportional to ${\bf k}\cdot{\bf p}\, p^2$ and 
 ${\bf k}\cdot{\bf p} \,p^i$ 
 can be dropped because they are subleading versus
 ${\bf p}_{1,2}\cdot{\bf k}\,p^2$ and 
 ${\bf k}\cdot{\bf p}_{1,2}\, p^i$ respectively. 
 The reason is that both ${\bf p}_{1}$, ${\bf p}_{2}$ and the final 
 momenta, which for $k \rightarrow 0$, can be approximated for 
 ${\bf p}_{1} + {\bf p}$ and   ${\bf p}_{2} - {\bf p}$, are close to 
 the Fermi surface, which means that $p \ll k_{F}$. 
 
 We still have to give the expressions for the polarization terms, 
 \begin{eqnarray}
 \epsilon_{0}({\bf k}) &=& \frac{\omega_{l}(k)}{k} \sqrt{Z_{l}(k)}\, , \\ 
 \epsilon_{j}({\bf k},\lambda) &=& \sqrt{Z_{t}(k)} e_{j}({\bf k},\lambda)\, ,
 \end{eqnarray}
 where $Z_{l}(k)$ and $Z_{t}(k)$ define the residue functions at the 
 poles in the external effective propagators \cite{neu} 
 and $e_{j}(\bf k,\lambda)$ 
 are orthogonal to $\bf k$ and normalized so that
 ${\bf e}({\bf k},\lambda) \cdot {\bf e}({\bf k},\lambda)^\ast = 1$.  
 To lowest order needed, 
 these are $Z_{l}= Z_{t} + {\cal O}(k^2)$. 
 The sum over transverse polatization vectors is 
 \begin{equation}
 \sum_{\lambda = \pm}e_{i}({\bf k},\lambda) e_{j}({\bf k},\lambda)^\ast = 
 \delta_{i j} - \hat{k}_{i} \hat{k}_{j} \, .
 \end{equation}  
 
 Now, it is convenient to make use of the fact that $\Gamma_{l,t}$ are 
 independent of the direction of ${\bf k}$. We can therefore average 
 the squares of ${\cal M}_{l,t}$ 
 over the directions of  ${\bf k}$. Then, the ${\bf p}_{1}$ and 
 ${\bf p}_{2}$ integrals can be done using the 
 `hard dense loop' results 
 \begin{eqnarray}
 \int \frac{d^3 p_{1}}{(2 \pi)^3}\, n_{F}({\bf p}_{1}) 
               (1-n_{F}({\bf p}_{1} + {\bf p})) 
               \delta\left(\omega - \frac{{\bf p}_{1}\cdot{\bf p}}{m} - 
                 \frac{p^2}{2  m}\right)  &=& \frac{m^2 \omega}
                       {4 \pi^2 p} \theta(\omega)
                        \theta\left(1- \frac{\omega^2}{v_{F}^2 p^2}
                        \right)\,, \\
 \int \frac{d^3 p_{1}}{(2 \pi)^3}\, p_{1}^i n_{F}({\bf p}_{1}) 
               (1-n_{F}({\bf p}_{1} + {\bf p})) 
               \delta\left(\omega - \frac{{\bf p}_{1}\cdot{\bf p}}{m} - 
                 \frac{p^2}{2  m}\right)  &=& \frac{m^3 \omega^2 p^i}
                       {4 \pi^2 p^3} \theta(\omega)
                       \theta\left(1- \frac{\omega^2}{v_{F}^2 p^2}
                       \right)\,, \\ 
 \int \frac{d^3 p_{1}}{(2 \pi)^3}\, p_{1}^2 n_{F}({\bf p}_{1}) 
               (1-n_{F}({\bf p}_{1} + {\bf p})) 
               \delta\left(\omega - \frac{{\bf p}_{1}\cdot{\bf p}}{m} - 
                 \frac{p^2}{2  m}\right)  &=& \frac{k_{F}^2 m^2 \omega}
                       {4 \pi^2 p} \theta(\omega)
                       \theta\left(1- \frac{\omega^2}{v_{F}^2 p^2}
                       \right)\,,
 \end{eqnarray}
 valid for $p \ll k_{F}$. Lastly, the spin and polarization summations 
 give 
 \begin{eqnarray}
  \Gamma_{l}(k) = \frac{g_{e}^6 m^2 k^2}{480 \pi^5 \Omega_p^5} 
                  \int_{0}^\infty & & dp \int_{0}^{v_{F} p} d\omega 
                  \int_{0}^{v_{F} p} d\omega^\prime 
                  \delta(\Omega_p - \omega - \omega^\prime) 
                  \nonumber \\  
         \times &[&  23\,\omega \omega^\prime p^4 \Omega_p^2 
                  |D(\omega,p)|^2 |D(\omega^\prime,p)|^2 \nonumber \\ 
          & &     + 4 \,\omega \omega^\prime (p^2 v_{F}^2 - \omega^2) 
                    |D(\omega^\prime,p)|^2 \nonumber \\ 
          & &        + 4 \,\omega \omega^\prime (p^2 v_{F}^2 - 
                        {\omega^\prime}^2) 
                     |D(\omega,p)|^2]\,,\\
   \Gamma_{t}(k) = \frac{g_{e}^6 m^2 k^2}{480 \pi^5 \Omega_p^5} 
                  \int_{0}^\infty & & dp \int_{0}^{v_{F} p} d\omega 
                  \int_{0}^{v_{F} p} d\omega^\prime 
                  \delta(\Omega_p - \omega - \omega^\prime) 
                  \nonumber \\  
         \times &[&  16\,\omega \omega^\prime p^4 \Omega_p^2 
                  |D(\omega,p)|^2 |D(\omega^\prime,p)|^2 \nonumber \\ 
          & &     + 3 \,\omega \omega^\prime (p^2 v_{F}^2 - \omega^2) 
                    |D(\omega^\prime,p)|^2 \nonumber \\ 
          & &        + 3 \,\omega \omega^\prime (p^2 v_{F}^2 - 
                        {\omega^\prime}^2) 
                     |D(\omega,p)|^2]\,.
 \end{eqnarray}  
 
 The integrals can be computed through the introduction of 
 the dimensionless variables $x = \omega/v_{F} p$, 
 $x = \omega^\prime/v_{F} p$ and $\overline{q} = v_{F} q/\Omega_{p}$. 
 The final results are 
 \begin{equation}\label{fin1}
 \Gamma_{l,t}(k) = a_{l,t} \left(\frac{\alpha}{v_{F}}\right)^{3/2} 
                          \frac{k^2}{200 m}\,                       
 \end{equation}
 where the constants are determined by numerical integration to be 
 \begin{eqnarray}
 a_{l}&\simeq&6.233\,87  \, , \label{fin2}\\
 a_{t}&\simeq&4.636\,08  \, .
 \label{fin3}
 \end{eqnarray}
 
 In conclusion, I have computed the leading contribution in 
 $v_{F}/c$ to 
 the damping rates of soft 
 quasiparticles in a degenerate non relativistic plasma for small 
 momentum $k$. This contribution comes from 
 density-density interactions in the plasma and  
 vanishes as $k \rightarrow 0$.  
 For $k = 0$, there are also contributions of higher orders in $v_{F}/c$ 
 coming from velocity 
 dependent interactions. These could  be computed within the same framework 
 by including one transverse effective interaction which would give 
 the subleading correction.     
 
 \subsection*{Acknowledgements}
 
 This work has been supported in part by funds provided by 
 CICYT, AEN96-1668.

\end{document}